\documentclass[twocolumn,trackchanges]{aastex62}

\newcommand{\kepler}{\textit{Kepler}}
\newcommand{\keplers}{\textit{Kepler's}}
\newcommand{\msun}{$M_{\odot}$}
\newcommand{\baseten}[1]{10$^{#1}$}

\newcommand{\rsun}{$R_{\odot}$}
\newcommand{\rhostar}{$\rho_{\star}$}

\newcommand{\qstar}{$Q'_{\star}$}
\newcommand{\micro}{$\mu$Hz}
\newcommand{\rstar}{$R_{\star}$}
\newcommand{\mstar}{$M_{\star}$}
\newcommand{\ms}{m s$^{-1}$}
\newcommand{\kms}{km s$^{-1}$}
\newcommand{\vsini}{$v\sin i$}
\newcommand{\ang}{\AA}
\newcommand{\rprs}{$R_{\text{P}}/R_{\star}$}
\newcommand{\kstar}{$K_{\star}$}
\newcommand{\occ}{$\delta_{\text{occ}}$}
\newcommand{\gcc}{g cm$^{-3}$}
\newcommand{\massp}{$M_{\rm p} $}
\newcommand{\massj}{$M_{\text{J}}$}
\newcommand{\radiusp}{$R_{\rm p} $}
\newcommand{\radiusj}{$R_{\text{J}}$}
\newcommand{\rhop}{$\rho_{\rm p}$}
\newcommand{\degrees}{$^{\text{o}}$}
\newcommand{\msy}{ms yr$^{-1}$}
\newcommand{\teff}{\mbox{$T_{\rm eff}$}}
\newcommand{\logg}{\mbox{$\log g$}}
\newcommand{\feh}{\mbox{$\rm{[m/H]}$}}
\newcommand{\Dnu}{\mbox{$\Delta \nu$}}
\newcommand{\dnu}{\mbox{$\Delta \nu$}}
\newcommand{\numax}{\mbox{$\nu_{\text{max}}$}}
\newcommand{\nuMax}{\mbox{$\nu_{\text{max}}$}}
\newcommand{\muHz}{\mbox{$\mu$Hz}}

\usepackage{amsmath}
\usepackage{graphicx}
\usepackage{tabularx}

\newcolumntype{L}[1]{>{\raggedright\arraybackslash}p{#1}}
\newcolumntype{C}[1]{>{\centering\arraybackslash}p{#1}}
\newcolumntype{R}[1]{>{\raggedleft\arraybackslash}p{#1}}

\graphicspath{{./}{figures/}}

\begin{document}

\title{The Curious Case of KOI 4: Confirming \keplers\ First Exoplanet Detection}

\correspondingauthor{Ashley Chontos}
\email{achontos@hawaii.edu}

\author[0000-0003-1125-2564]{Ashley Chontos}
\affiliation{Institute for Astronomy, University of Hawai`i, 2680 Woodlawn Drive, Honolulu, HI 96822, USA}
\affiliation{NSF Graduate Research Fellow}

\author[0000-0001-8832-4488]{Daniel Huber}
\affiliation{Institute for Astronomy, University of Hawai`i, 2680 Woodlawn Drive, Honolulu, HI 96822, USA}

\author[0000-0001-9911-7388]{David W. Latham}
\affiliation{Harvard-Smithsonian Center for Astrophysics, 60 Garden Street, Cambridge, MA 02138, USA}

\author[0000-0001-6637-5401]{Allyson Bieryla}
\affiliation{Harvard-Smithsonian Center for Astrophysics, 60 Garden Street, Cambridge, MA 02138, USA}

\author{Vincent Van Eylen}
\affiliation{Department of Astrophysical Sciences, Princeton University, 4 Ivy Lane, Princeton, NJ 08540, USA}
\affiliation{Leiden Observatory, Leiden University, 2333CA Leiden, The Netherlands}

\author{Timothy R. Bedding}
\affiliation{Sydney Institute for Astronomy (SIfA), School of Physics, University of Sydney, NSW 2006, Australia}
\affiliation{Stellar Astrophysics Centre, Department of Physics and Astronomy, Aarhus University, Ny Munkegade 120, DK-8000, Aarhus C, Denmark}

\author[0000-0002-2580-3614]{Travis Berger}
\affiliation{Institute for Astronomy, University of Hawai`i, 2680 Woodlawn Drive, Honolulu, HI 96822, USA}

\author[0000-0003-1605-5666]{Lars A. Buchhave}
\affiliation{DTU Space, National Space Institute, Technical University of Denmark, Elektrovej 328, DK-2800 Kgs. Lyngby, Denmark}

\author[0000-0002-4588-5389]{Tiago L. Campante}
\affiliation{Instituto de Astrof\'{\i}sica e Ci\^{e}ncias do Espa\c{c}o, Universidade do Porto, Rua das Estrelas, PT4150-762 Porto, Portugal}
\affiliation{Departamento de F\'{\i}sica e Astronomia, Faculdade de Ci\^{e}ncias da Universidade do Porto, Rua do Campo Alegre, s/n, PT4169-007 Porto, Portugal}

\author[0000-0002-5714-8618]{William J. Chaplin}
\affiliation{School of Physics and Astronomy, University of Birmingham, Edgbaston, Birmingham, B15 2TT, UK}
\affiliation{Stellar Astrophysics Centre, Department of Physics and Astronomy, Aarhus University, Ny Munkegade 120, DK-8000, Aarhus C, Denmark}

\author[0000-0001-8196-516X]{Isabel L. Colman}
\affiliation{Sydney Institute for Astronomy (SIfA), School of Physics, University of Sydney, NSW 2006, Australia}
\affiliation{Stellar Astrophysics Centre, Department of Physics and Astronomy, Aarhus University, Ny Munkegade 120, DK-8000, Aarhus C, Denmark}

\author[0000-0003-1634-9672]{Jeff L. Coughlin}
\affiliation{NASA Ames Research Center, Moffett Field, CA 94035, USA}
\affiliation{SETI Institute, 189 Bernardo Avenue, Suite 200, Mountain View, CA 94043, USA}

\author[0000-0002-4290-7351]{Guy Davies}
\affiliation{School of Physics and Astronomy, University of Birmingham, Edgbaston, Birmingham, B15 2TT, UK}
\affiliation{Stellar Astrophysics Centre, Department of Physics and Astronomy, Aarhus University, Ny Munkegade 120, DK-8000, Aarhus C, Denmark}

\author{Teruyuki Hirano}
\affiliation{Department of Earth and Planetary Sciences, Tokyo Institute of Technology, 2-12-1 Ookayama, Meguro-ku, Tokyo 152-8551, Japan}
\affiliation{Institute for Astronomy, University of Hawai`i, 2680 Woodlawn Drive, Honolulu, HI 96822, USA}

\author{Andrew W. Howard}
\affiliation{Department of Astronomy, California Institute of Technology, 1200 E California Boulevard, Pasadena, CA 91125, USA}

\author{Howard Isaacson}
\affiliation{Astronomy Department, University of California, Berkeley, CA 94720, USA}

\begin{abstract}

\noindent
The discovery of thousands of planetary systems by \kepler\ has demonstrated that planets are ubiquitous. However, a major challenge has been the confirmation of \kepler\ planet candidates, many of which still await confirmation. One of the most enigmatic examples is KOI 4.01, \keplers\ first discovered planet candidate detection (as KOI 1.01, 2.01, and 3.01 were known prior to launch). Here we present the confirmation and characterization of KOI 4.01 (now Kepler-1658), using a combination of asteroseismology and radial velocities. Kepler-1658 is a massive, evolved subgiant (\mstar\ = 1.45 $\pm$ 0.06 \msun, \rstar\ = 2.89 $\pm$ 0.12 \rsun) hosting a massive (\massp\ = 5.88 $\pm$ 0.47 \massj, \radiusp\ = 1.07 $\pm$ 0.05 \radiusj) hot Jupiter that orbits every 3.85 days. Kepler-1658 joins a small population of evolved hosts with short-period ($\le$100 days) planets and is now the closest known planet in terms of orbital period to an evolved star. Because of its uniqueness and short orbital period, Kepler-1658 is a new benchmark system for testing tidal dissipation and hot Jupiter formation theories. Using all 4 years of \kepler\ data, we constrain the orbital decay rate to be $\dot{P}$ $\le$ -0.42 s yr$^{-1}$, corresponding to a strong observational limit of \qstar\ $\ge$ 4.826 $\times$ \baseten{3} for the tidal quality factor in evolved stars. With an effective temperature \teff\ $\sim$6200 K, Kepler-1658 sits close to the spin-orbit misalignment boundary at $\sim$6250 K, making it a prime target for follow-up observations to better constrain its obliquity and to provide insight into theories for hot Jupiter formation and migration.

\end{abstract}


\section{Introduction}

The \kepler\ mission \citep{borucki2010,kock2010,borucki2016} revolutionized the field of exoplanetary science. Pre-\kepler\ exoplanet discoveries were biased towards close-in giant planets (``hot Jupiters''), a planet type absent from our own solar system. However, \kepler\ later revealed that hot Jupiters are in fact rare, and smaller sub-Neptune sized planets are ubiquitous in inner planetary systems \citep{howard2012,dressing2013,petigura2013,gaidos2014,morton2014,silburt2015}.

When the \kepler\ spacecraft launched in March 2009, three planets in the \kepler\ field were already known from ground-based transit observations \citep{odonovan2006,pal2008,bakos2010}. These targets were designated the first three KOI (\kepler\ Object of Interest) numbers, making KOI 4.01 \keplers\ first new planet candidate (PC). The initial classification in the Kepler Input Catalog \citep[KIC,][]{brown2011} for KOI 4 implied a 1.1 solar radius (\rsun) main-sequence star with an effective temperature (\teff) of 6240 K \citep{brown2011}. Based on a primary transit depth of 0.13\%, this stellar classification implied that KOI 4 is orbited by a Neptune-sized planet. However, because a deep secondary eclipse was observed, KOI 4.01 was marked as a false positive (FP) in early \kepler\ KOI catalogs, since a secondary eclipse would not be observable for a Neptune-sized planet orbiting a main sequence star. 

The NASA Exoplanet Archive reveals a more detailed picture of the complex vetting history of Kepler's first exoplanet candidate. KOI 4.01 was not listed in the first KOI catalog \citep{borucki2011a} but appeared as a `moderate probability candidate' in the second KOI catalog \citep{borucki2011}, with the host star noted as a rapid rotator (\vsini\ = 40 \kms). In the third catalog, \citet{batalha2013} listed KOI 4.01 as a PC but it was marked back to a FP in the fourth catalog \citep{burke2014}, likely due to the secondary eclipse. The fifth \citep{rowe2015} and sixth \citep{mullally2015} catalogs did not disposition existing KOIs within certain parameter spaces.
%In the last two \kepler\ data releases, planet candidate vetting was automated via the Robovetter pipeline \citep{coughlin2016}. 
The seventh catalog \citep{coughlin2016} was the first fully uniform catalog using the Robovetter pipeline, marking 237 KOIs that were previously FPs back to PCs using updated stellar parameters, including KOI 4. In the final catalog \citep{thompson2018}, the Robovetter also dispositioned it as a PC. Until now, \keplers\ first new planet candidate has awaited confirmation as a genuine planet detection.

\begin{figure}
\includegraphics[width=\linewidth]{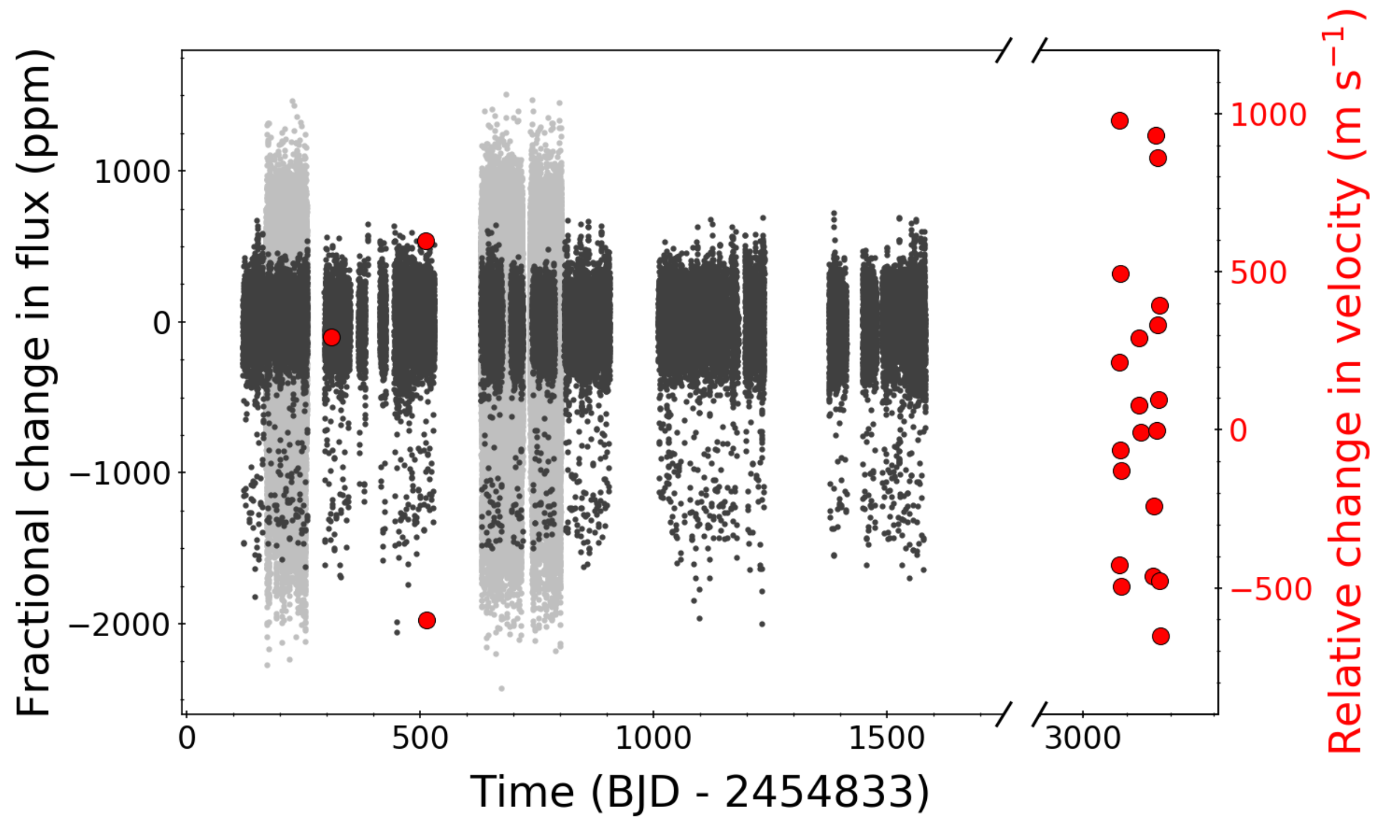}
\caption{\kepler\ photometry and radial velocity observations for KOI 4.01. Long-cadence photometric data are shown in dark gray while short-cadence are shown in light gray. Three high-resolution spectra (red points) were initially taken of KOI 4.01 during the mission before it was marked as a false positive. This was followed by a break of 7 years before it was re-observed by our team in 2017.}
\label{fig:combined}
\end{figure}

Systems like KOI 4.01 are interesting because giant planets at short orbital periods ($P < 100$~days) are rare around subgiant stars \citep[e.g.][]{johnson2007,johnson2010,reffert2015,lillo-box2016,veras2016}, although the reason for this is still a topic of debate. On one hand, this may be related to the stellar mass. Subgiant host stars are thought to be more massive than main sequence stars targeted for planet detection. A higher mass could shorten the lifetime of the protoplanetary disk and lead to fewer short-period giant planets orbiting these type of stars \citep[e.g.][]{burkert2007,kretke2009}. Other authors have suggested that subgiants have fewer short-period planets because these objects may get destroyed by tidal evolution, which is likely stronger for more evolved stars \citep[e.g.][]{villaver2009,schlaufman2013}. Distinguishing between those scenarios is further complicated by the fact that it is challenging to derive stellar masses of evolved stars \citep{lloyd2011,lloyd2013,johnson2013,ghezzi2018}.

Here we confirm and characterize KOI 4.01, hereafter Kepler-1658, using a combination of asteroseismology and spectroscopic follow-up observations. Due to its short orbital period, Kepler-1658 b is an ideal target to constrain the role of tides around more evolved stars. In addition, we are able to constrain the stellar mass and other stellar parameters to high precision and accuracy by analyzing the stellar oscillations and comparing these to stellar models. We conclude by discussing future observations that could provide insight for theories of hot Jupiter formation and migration.

\section{Observations}

\subsection{\kepler\ Photometry}

\begin{table*}[t]
\caption{TRES Radial Velocity Observations}
\renewcommand{\tabcolsep}{6mm}
\center
\begin{tabular}{lccccc}
\noalign{\smallskip}
\hline\hline
\noalign{\smallskip}
Time (BJD TDB) & Phase$^1$ & RV (\ms) &  $\sigma_{\text{RV}}$ (\ms) & BS$^2$ (\ms) & $\sigma_{\text{BS}}$ (\ms) \\
\noalign{\smallskip}
\hline
\noalign{\smallskip}
2455143.566982 & 39.757 & -648.64 & 261.20 & -11.1 & 205.7 \\
2455343.873508 & 91.793 & -344.33 & 223.72 & -22.9 & 138.1 \\
2455345.873597 & 92.313 & -1542.39 & 153.02 & 244.2 & 72.1 \\
2457914.843718 & 759.687 & 0.00 & 155.26 & -114.2 & 41.6 \\
2457915.936919 & 759.971 & -756.57 & 165.32 & -122.5 & 64.7 \\
2457916.865409 & 760.212 & -1391.09 & 159.21 & 220.9 & 58.9 \\
2457917.929198 & 760.488 & -1031.39 & 103.84 & -5.9 & 70.9 \\
2457918.888799 & 760.738 & -480.41 & 155.26 & -87.2 & 52.9 \\
2457919.869505 & 760.992 & -1095.83 & 109.08 & 108.9 & 59.6 \\
2457920.859134 & 761.250 & -1458.12 & 187.52 & 238.5 & 95.2 \\
2457960.932038 & 771.660 & -679.80 & 156.99 & -40.2 & 136.6 \\
2457961.930294 & 771.919 & -891.62 & 232.50 & 26.1 & 127.6 \\
2457965.808860 & 772.927 & -977.16 & 131.20 & -142.2 & 61.5 \\
2457993.735605 & 780.182 & -1424.46 & 200.28 & 21.1 & 96.0 \\
2457994.753312 & 780.446 & -1205.20 & 190.75 & 168.0 & 50.2 \\
2457999.668199 & 781.723 & -48.17 & 137.98 & -223.2 & 68.9 \\
2458001.720710 & 782.256 & -971.08 & 146.58 & -48.9 & 89.2 \\
2458002.718939 & 782.515 & -640.20 & 127.69 & -98.9 & 60.0 \\
2458003.661261 & 782.760 & -118.49 & 103.28 & -274.6 & 85.1 \\
2458006.701108 & 783.550 & -873.64 & 137.11 & 66.8 & 77.7 \\
2458007.747359 & 783.822 & -578.74 & 130.97 & -45.6 & 66.9 \\
2458008.800582 & 784.095 & -1439.71 & 156.00 & -6.5 & 74.5 \\
2458009.717813 & 784.333 & -1611.87 & 145.55 & 149.4 & 40.9 \\
\noalign{\smallskip}
\hline
\noalign{\smallskip}
\multicolumn{6}{l}{{\sc \textbf{Notes ---}}} \\
\multicolumn{6}{l}{$^1$ Indicates the orbital phase of the planet at the time of observation (where 0 phase is defined as the start of \kepler).}\\
\multicolumn{6}{l}{$^2$ Line bisector spans and  uncertainties, as discussed in Section 2.3.}\\
\end{tabular}
\label{tab:rvs} 
\end{table*}

The \kepler\ spacecraft had two observing modes: long-cadence \citep[29.4 min;][]{jenkins2010} and short-cadence \citep[58.85 s;][]{gilliland2010sc}. In the nominal \kepler\ mission most \kepler\ targets were observed in long-cadence, while 512 short-cadence slots remained for select targets. Short-cadence observations are important for asteroseismology of dwarfs and subgiants, whose oscillations occur on timescales faster than 30 minutes \citep{gilliland2010astero,chaplin2013astero}. 

Decisions on which targets to observe in short-cadence were made on a quarter-by-quarter basis. In particular, once planet candidates were detected and assigned a KOI number, targets were put on short-cadence if the probability of detecting oscillations was deemed significant \citep{chaplin2011probs}. Kepler-1658 was observed in long-cadence for most of the mission aside from 3 quarters, while it was only observed in short-cadence in Quarters 2, 4, 7, and 8, for a total of 213.7 days (Figure \ref{fig:combined}).

\subsection{Imaging}
\label{sec:imaging}

Kepler-1658 was observed by Robo-AO, a robotic, visible light, laser adaptive optics (AO) imager that searched for nearby companions which could potentially contaminate target light curves \citep{law14,baranec2016,ziegler17}. \citet{law14} reported a nearby companion to Kepler-1658 at a separation of 3.42" and a contrast of 4.46 mag in the LP600 filter, which has a similar wavelength coverage to the \kepler\ bandpass. In addition to Robo-AO, the Kepler UKIRT survey reported a detection of the same companion with a contrast of 4.23 $\Delta$mag in the J-band \citep{furlan2017}. Gaia Data Release 2 reported parallaxes of 1.24 $\pm$ 0.03 mas and 0.75 $\pm$ 0.05 mas corresponding to Kepler-1658 and its companion, respectively \citep{lindegren2016}. Therefore, we conclude that the two targets are not physically associated.

\subsection{Spectroscopy and Radial Velocities}

Initial spectroscopic follow-up of Kepler-1658 was obtained by the Kepler Follow-up Observing Program (KFOP), including the HIRES spectrograph \citep{vogt1994} on the 10-m telescope at Keck Observatory (Mauna Kea, Hawaii), the FIES spectrograph \citep{djupvik2010} on the 2.5-m Nordic Optical Telescope at the Roque de los Muchachos Observatory (La Palma, Spain), and the Tillinghast Reflector Echelle Spectrograph (TRES) \citep{furesz08} on the 1.5-m Tillinghast reflector at the F. L. Whipple Observatory (Mt. Hopkins, Arizona). The observing notes archived at the Kepler Community Follow-up Observing Program (CFOP)\footnote{\url{https://exofop.ipac.caltech.edu/kepler/}} show that these spectra confirmed that Kepler-1658 is a rapid rotator which, combined with the detection of the close companion (see previous section), discouraged further follow-up observations to confirm the planet candidate.

Following the asteroseismic reclassification of the host star (see next section), we initiated an intensive radial-velocity follow-up program using TRES, a fiber-fed \'echelle spectrograph spanning the spectral range 3900-9100 \AA ngstroms with a resolving power of R$\sim$44,000. We obtained 23 spectra with TRES between UT November 08 2009 and September 13 2017 using the medium 2.3" fiber. The spectra were reduced and extracted as outlined in \citet{buchhave2010}. The average exposure time of $\sim$1800 seconds, corresponding to a mean signal-to-noise (S/N) per resolution element of $\sim$53 at the peak of the continuum near the Mg b triplet at 519nm. We used the strongest S/N spectrum as a template to derive relative radial velocities by cross-correlating the remaining spectra order-by-order against the template, which is given a relative velocity of 0 \kms, by definition.

Monitoring of standard stars with TRES shows that the long-term zero point of the instrument is  stable to within $\pm$ 5 \ms\ over recent years. Due to mechanical and optical upgrades to TRES in the early years, there were major shifts in the instrumental zero point of the velocity system. The correction for the 2009 observation was -115 \ms\ and the correction for both 2010 observations was -82 \ms.

\begin{table}[htb]
\begin{centering}
\caption{Stellar Parameters}
\renewcommand{\tabcolsep}{0mm}
\begin{tabular}{l c}
\noalign{\smallskip}
\tableline\tableline
\noalign{\smallskip}
\textbf{Parameter} &  \textbf{KIC 3861595} \\
\noalign{\smallskip}
\tableline\tableline
\noalign{\smallskip}
\multicolumn{2}{c}{Basic Properties} \\
\noalign{\smallskip}
\hline
\noalign{\smallskip}
2MASS ID & 19372557+3856505 \\
Right Ascension	& 19 37 25.575 \\
Declination	& +38 56 50.515 \\
Magnitude (\textit{Kepler})	& 10.195 \\
Magnitude ($V$) & 11.62 \\
Magnitude (TESS) & 10.98  \\
\noalign{\smallskip}
\hline
\noalign{\smallskip}
\multicolumn{2}{c}{Spectroscopy} \\
\noalign{\smallskip}
\hline
\noalign{\smallskip}
Effective Temperature, \teff\ (K) & $6216 \pm 78$ \\
Metallicity, [m/H] & $-0.18 \pm 0.10$ \\
Projected rotation speed, \vsini\ (\kms)$^{\rm *}$ & $33.95 \pm 0.97$ \\
\noalign{\smallskip}
\hline
\noalign{\smallskip}
\multicolumn{2}{c}{Asteroseismology} \\
\noalign{\smallskip}
\hline
\noalign{\smallskip}
Stellar Mass, \mstar\ (\msun)& $1.447 \pm 0.058$\\
Stellar Radius, \rstar\ (\rsun)& $2.891^{+0.130}_{-0.106}$ \\
Stellar Density, \rhostar\ (\gcc)& $0.0834 \pm 0.0079$ \\
Surface Gravity, \logg\ (dex) & $3.673 \pm 0.026 $ \\
\noalign{\smallskip}
\hline
\noalign{\smallskip}
\multicolumn{2}{l}{$^{\rm *}$ Using FIES spectrum, discussed in Section 3.3.}\\
\end{tabular}
\end{centering}
\label{tab:stellar}
\end{table}

A bisector analysis was performed on the TRES spectra as described in \citet{torres2007} to check for asymmetries in the line profile which could be indicative of an unresolved eclipsing binary. The line bisector spans (BS) showed no correlation with the measured radial velocities and are small compared to the orbital semi-amplitude. All relative velocities, bisector values, and associated uncertainties are listed in Table \ref{tab:rvs}.

\section{Host Star Characterization}

\subsection{Atmospheric Parameters}

Atmospheric parameters were derived from the TRES and FIES spectra using the Stellar Parameter Classification code 
\citep[SPC, see][]{buchhave2012metals}. We adopted a weighted mean of the solutions to the individual spectra, yielding $\teff=6216\pm51$\,K, $\logg=3.57\pm0.1$\,dex and $\feh=-0.18\pm0.08$\,dex. The SPC-derived \logg\ is in good agreement with the asteroseismic detection (see below), and thus no iterations between the spectroscopic and asteroseismic solution were required. 

We also analyzed the HIRES spectrum using Specmatch-emp \citep{yee2017}, yielding consistent values within 2$\sigma$ ($\teff=6241\pm110$\,K, $\feh=-0.05\pm0.09$\,\,dex). We adopted the weighted SPC values as our final solution, and added 59\,K in \teff\ and 0.062\,dex in \feh\ in quadrature to the formal uncertainties to account for systematic differences between spectroscopic methods \citep{torres2012}. The final adopted values are listed in Table \ref{tab:stellar}.

\subsection{Asteroseismology}

Asteroseismology, the study of stellar oscillations, and transits form a powerful synergy to investigate exoplanet systems.
\citep{stello2009astero,gilliland2010astero,huber2013astero,vaneylen2014,huber2015exo}. Currently, more than one hundred \kepler\ exoplanet host stars have been characterized through asteroseismology \citep{huber2013astero,lundkvist2016}.

\begin{figure}
\includegraphics[width=\linewidth]{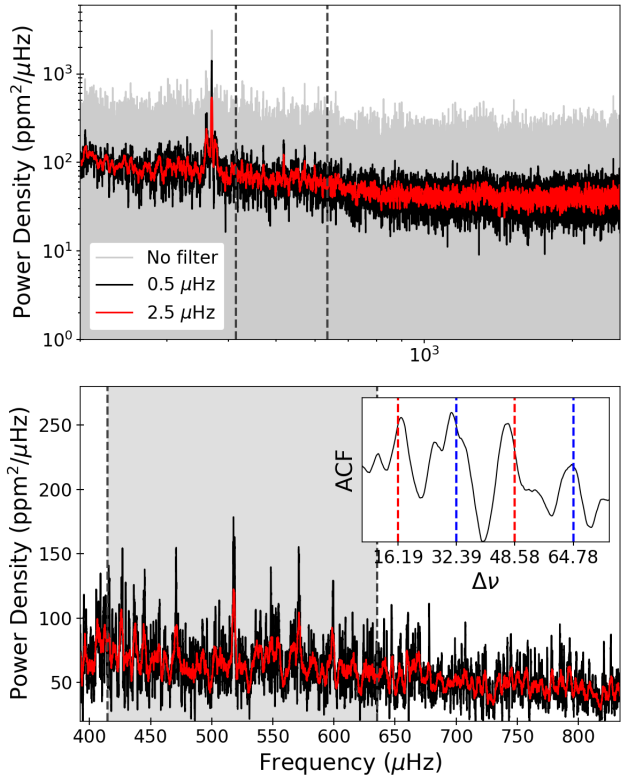}
\caption{Power spectrum of the \kepler\ short-cadence data for Kepler-1658. Top: Power spectrum in log-log scale, where the region of oscillations is marked by the dashed lines. The original power is shown in gray and smoothing filters of widths 0.5 and 2.5 \micro\ are shown in black and red, respectively. Bottom: Power spectrum in linear space zoomed in on the region of oscillations. The shaded area highlights the power that is used to calculate the autocorrelation, shown in the inset.}
\label{fig:PowerSpectrum}
\end{figure}

We performed a search for oscillations in Kepler-1658 in the \kepler\ short-cadence data. Asteroseismic analysis included removing any data with nonzero quality flags, clipping transits and outliers, then normalizing each individual quarter before concatenating the light curve. A high-pass filter was used to remove long-period systematics before computing the power spectrum. Box filters of widths 0.5 and 2.5 \muHz\ were used to smooth the power spectrum in order to make the signal clear. The power spectrum for Kepler-1658 can be seen in Figure \ref{fig:PowerSpectrum}, showing a characteristic frequency-dependent noise due to granulation and a power excess marked by the gray dashed lines. We note that the strong peak near $\sim$300 \muHz\ is a well-known artefact of \kepler\ short-cadence data \citep{gilliland2010sc}.

\begin{figure}
\includegraphics[width=\linewidth]{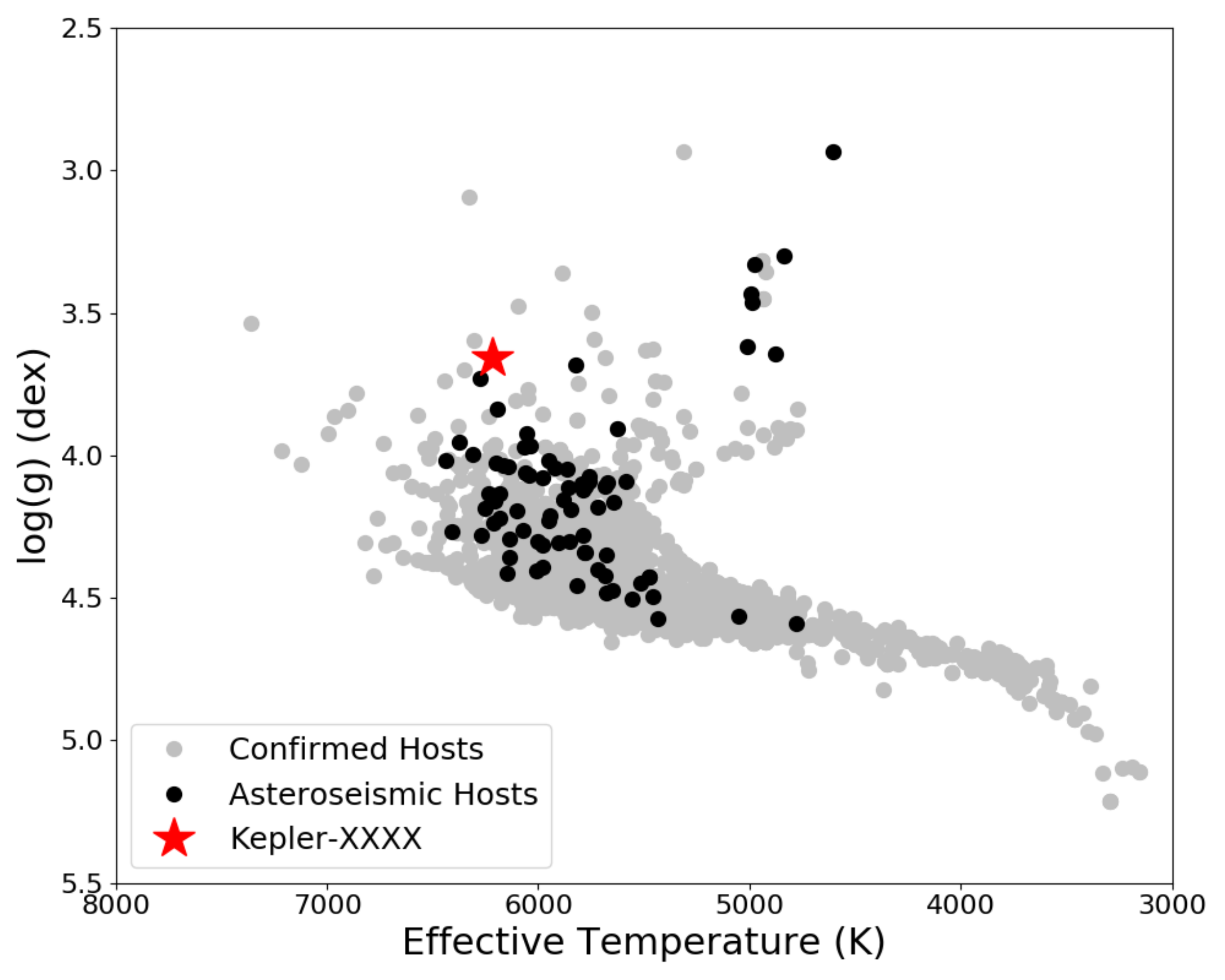}
\caption{Surface gravity versus effection temperature for confirmed \kepler\ exoplanet hosts. Gray points represent confirmed hosts, with known asteroseismic hosts in black. Kepler-1658, represented by the red star, sits in an underpopulated area of stellar parameter space as a massive, evolved subgiant.}\
\label{fig:HRdiagram}
\end{figure}

Since the power excess has relatively low S/N, we used an autocorrelation to confirm the oscillations. The region with excess power should have a width that we can estimate by using a linear scaling relation:
\begin{equation}
w = w_{\odot} \bigg(\frac{\nu_{\text{max}}}{\nu_{\text{max}, \odot}}\bigg),
\end{equation}
where $w_{\odot}$ = 1300 \muHz\ is the width of oscillations in the Sun. To prevent adding noise to the calculated autocorrelation, only the power in this region was used. Confirming the oscillations requires detecting peaks with a regular spacing that follows the well-known correlation between \dnu\ and \numax\ \citep{stello2009scaling,hekker2009,mosser2010,hekker2011a,hekker2011b,huber2011}. The autocorrelation of the power spectrum is shown in the inset in Figure \ref{fig:PowerSpectrum} and confirms the detection of oscillations. Red and blue dashed lines mark the expected positions of regular spacings based on asteroseismic scaling relations. Red is the expected spacing of adjacent radial and dipole modes ($\sim\!n$\dnu$/2$) and blue is the expected spacing of consecutive radial modes ($\sim\!n$\dnu).

The low S/N of the seismic detection, combined with the possible presence of mixed modes, make the autocorrelation an imprecise tool to measure $\Dnu$. Additionally, the short-cadence artefact at $\approx 300\,\muHz$ prevents a reliable background fit to the power spectrum and thus a measurement of \nuMax. To determine \Dnu, we computed \'echelle diagrams over a grid of trial $\Dnu$ values separated by 0.1\,\muHz\ to identify the spacing which produces straight ridges of modes of consecutive overtones, a method commonly adopted to verify the accuracy of $\Dnu$ values \citep{bedding10}. We identified 32.5\,$\muHz$ as the correct spacing, consistent with four independent analyses by co-authors \citep{campante2010,chaplin2011,bedding2012,davies2016}. We adopted the result from the manual analysis as our final value and adopted the scatter over all methods as an uncertainty, yielding $\Dnu=32.5\pm1.6\,\muHz$.

\begin{figure}
\includegraphics[width=\linewidth]{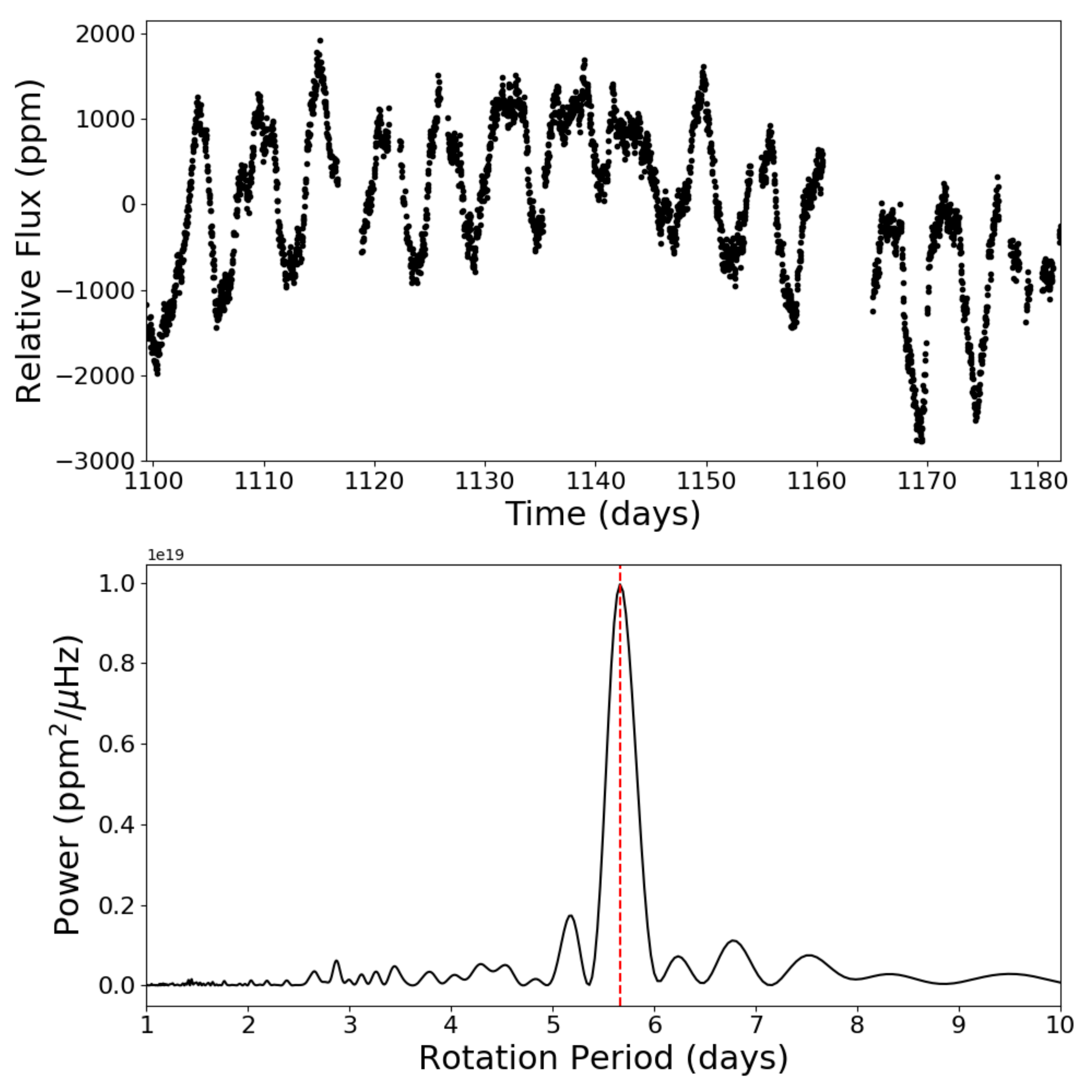}
\caption{Top: Transit-clipped Quarter 11 long-cadence light curve for Kepler-1658. Bottom: Lomb-Scargle periodogram showing a strong peak at 5.66 $\pm$ 0.31 days, which we interpret as the stellar rotation period.}
\label{fig:rotation}
\end{figure}

Since the low S/N does not allow a reliable constraints on individual frequencies or \numax, we used grid-based modeling \citep{gai2011} with atmospheric parameters from spectroscopy and the asteroseismic $\Dnu$ to derive a full set of host star properties. To perform grid-modeling we used the open-source code \texttt{isoclassify}\footnote[1]{\url{https://github.com/danxhuber/isoclassify}} \citep{huber2017}, which adopts a grid of MIST isochrones \citep{choi2016} to probabilistically infer stellar parameters given any combination of photometric, spectroscopic or asteroseismic input parameters and adopts theoretically motivated corrections for the $\Dnu$ scaling relation from \citet{sharma2016}. The results confirm that Kepler-1658 is a relatively massive (\mstar\ = 1.45 $\pm$ 0.06 \msun) and evolved (\rstar\ = 2.89 $\pm$ 0.12 \rsun) subgiant star (Table \ref{tab:stellar}). Kepler-1658 joins a small sample of subgiant host stars for which the stellar mass is accurately determined through asteroseismology (Figure \ref{fig:HRdiagram}).

We note that $\Dnu$ scaling relation is based on simplified assumptions compared to analyses using individual mode frequencies, and thus may be affected by systematic errors. However, independent model calculations have demonstrated that the relation is accurate to $<1$\% in $\Dnu$ ($<$0.5\% in $\rho$) for stars in the \teff\ and \feh\ range of Kepler-1658 \citep{white11,rodrigues17}. Therefore, any potential systematic error in \rhostar\ introduced by using the $\Dnu$ scaling relation is negligible compared to our adopted uncertainties.

\subsection{Stellar Rotation}

The top panel of Figure \ref{fig:rotation} shows the unfiltered light curve of Kepler-1658 from Quarter 11, demonstrating strong evidence for rotational modulation due to spots. The photometric variability has an amplitude of $\sim$0.1\% and shows a strong peak at 5.66 $\pm$ 0.31 days in the Lomb-Scargle (LS) periodogram (bottom panel of Figure \ref{fig:rotation}). We tested for temporal variations of the stellar rotation by calculating a LS periodogram of the unfiltered light curve for each quarter available. The analysis demonstrated that the equatorial rotation velocity does not change over the \kepler\ baseline. Combining this rotation period with the asteroseismic radius,  we compute an equatorial rotation velocity, $v$ = 25.82 $\pm$ 1.77 \kms.

\begin{figure}
\includegraphics[width=\linewidth]{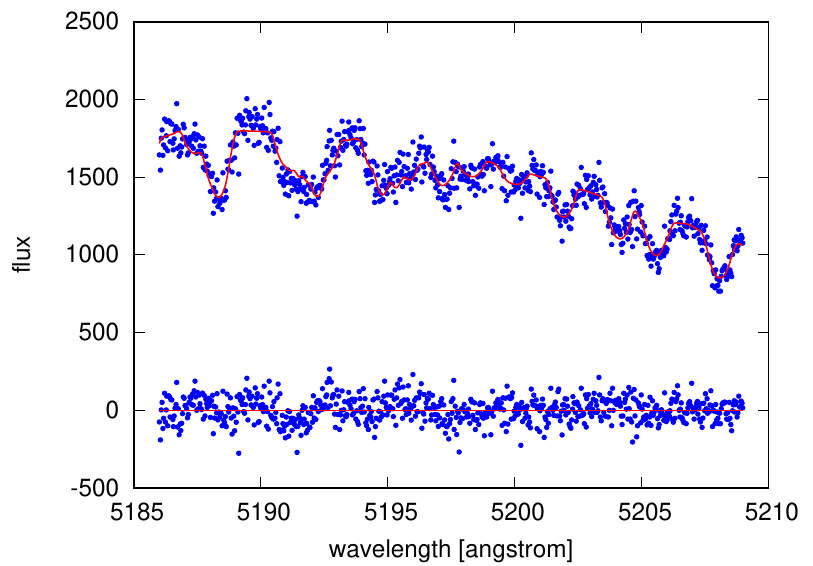}
\caption{One FIES spectral segment of Kepler-1658. A theoretical, unbroadened spectrum is convolved with rotation and a macroturbulent broadening kernel to fit the spectrum and is shown in red, where the scatter in the residuals of the best-fit \vsini\ value is shown below the spectrum.}\
\label{fig:vsini}
\end{figure}

In order to estimate the projected rotation velocity (\vsini) of Kepler-1658, we analyzed the FIES spectrum using the technique described by \citet{hirano2012}. In brief, we convolved a theoretical, unbroadened spectrum generated by adopting the stellar parameters for Kepler-1658 \citep{coelho2005} with the rotation plus macroturbulence broadening kernel (and instrumental profile), assuming the radial-tangential model \citep{gray2005}. The broadening kernel has several parameters, including \vsini, the macroturbulent velocity $\zeta$, and stellar limb-darkening parameters, but we only optimized \vsini, along with the overall normalization parameters describing the spectrum continuum. 

\begin{figure}
\includegraphics[width=\linewidth]{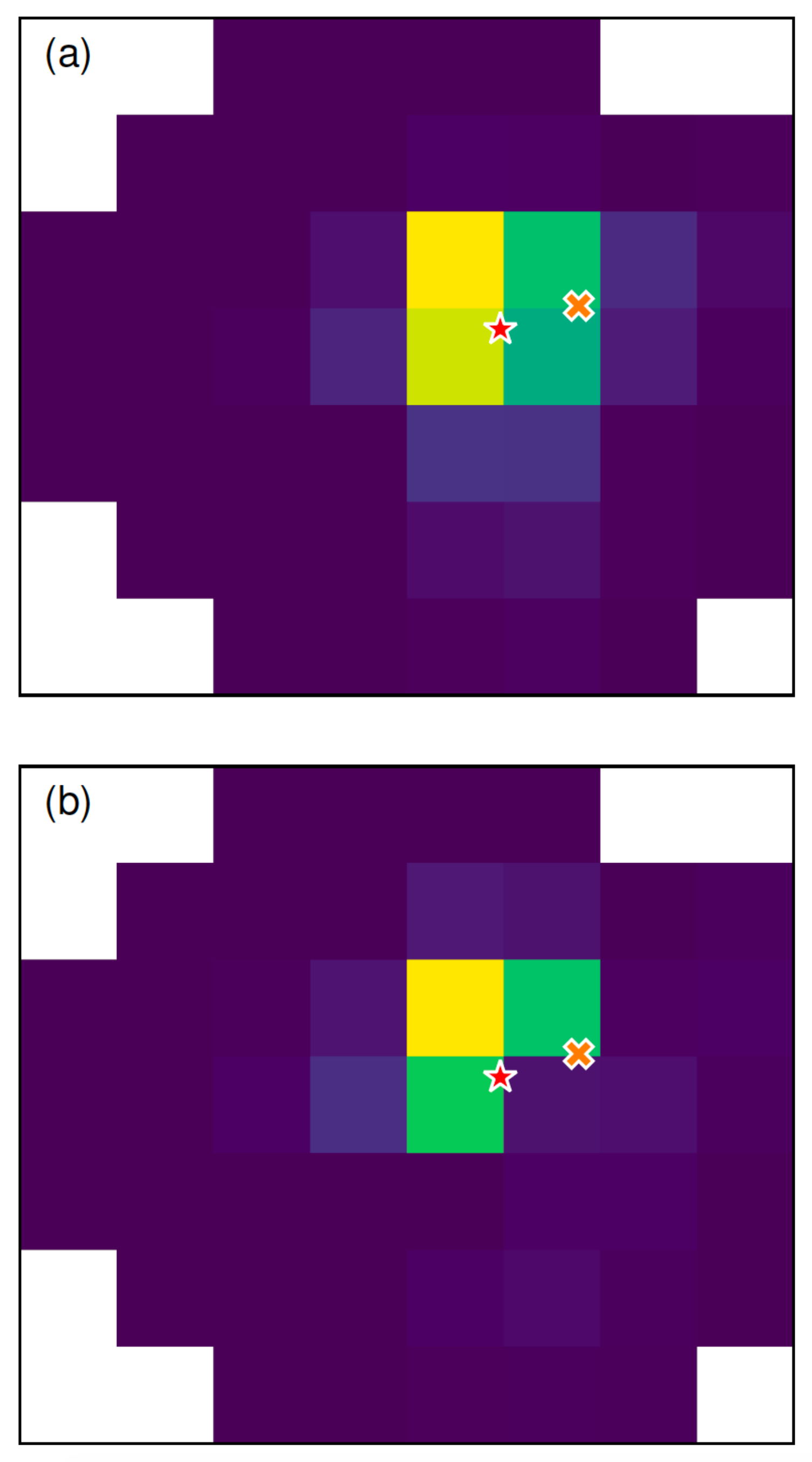}
\caption{Panel (a): Target pixel files of Kepler-1658 averaged over one full quarter. Panel (b): A difference image using frames coinciding with the maxima and minima of a phase curve calculated from the measured rotation period. The star marks the location of Kepler-1658, and the companion identified using AO imaging is marked with a cross.}
\label{fig:difference}
\end{figure}

We attempted the fits for three different spectral segments (5126-5154 \ang, 5186-5209 \ang, 5376-5407 \ang), where an example of one segment is shown in Figure \ref{fig:vsini}. The uncertainty of \vsini\ was derived based on the scatter of the best-fit values for these segments. For the macroturbulent velocity, we adopted $\zeta=4.7\pm 1.2$ \kms\ based on the empirical relation 
between \teff\ and $\zeta$ derived by \citet{hirano2014}, but we found that choice of $\zeta$ has very little impact on the estimated \vsini, due to the latter's large value (variation less than $0.2$ \kms). We found the \vsini\ for Kepler-1658 to be $33.95\pm 0.97$ \kms. 

There is a clear discrepancy between the equatorial rotation velocity ($v$) and projected rotation velocity (\vsini) we computed, with \vsini\ $\gg$ $v$. As discussed in Section 2.2, there is a companion within the 4" \kepler\ pixel and is therefore not resolved. As a result, we tested whether this rotational signal is coming from Kepler-1658 or from its neighbor using \kepler\ target pixel files (TPFs).

Since the angular separation between Kepler-1658 and its neighbor is the size of a \kepler\ pixel, we wanted to identify in which pixel the rotational signal is strongest. We used the difference imaging technique \citep{bryson2013,colman2017}, which involved phasing the light curve on the period of the rotational signal, binning by a factor of 1000, and selecting the data that fell within $\pm$ 1\% of the peaks and troughs of the phased light curve. To create the difference image, we subtracted the data around the troughs from the data around the peaks. We did this for each pixel, creating a difference image which gives an indication of the relative strength of the rotational signal over the postage stamp. We then compared the difference image to an average image from the same observing quarter (Figure \ref{fig:difference}), and found that in 11 of the 17 quarters the pixel with the brightest flux is the same as the pixel where the rotational signal is the strongest. Differences in the other 6 quarters did not exceed one pixel and were inconsistent with the relative positions of the KOI-4 and the imaged companion (see symbols in Figure 5). These results strongly imply that the rotational signal is coming from Kepler-1658.

\begin{figure}
\includegraphics[width=\linewidth]{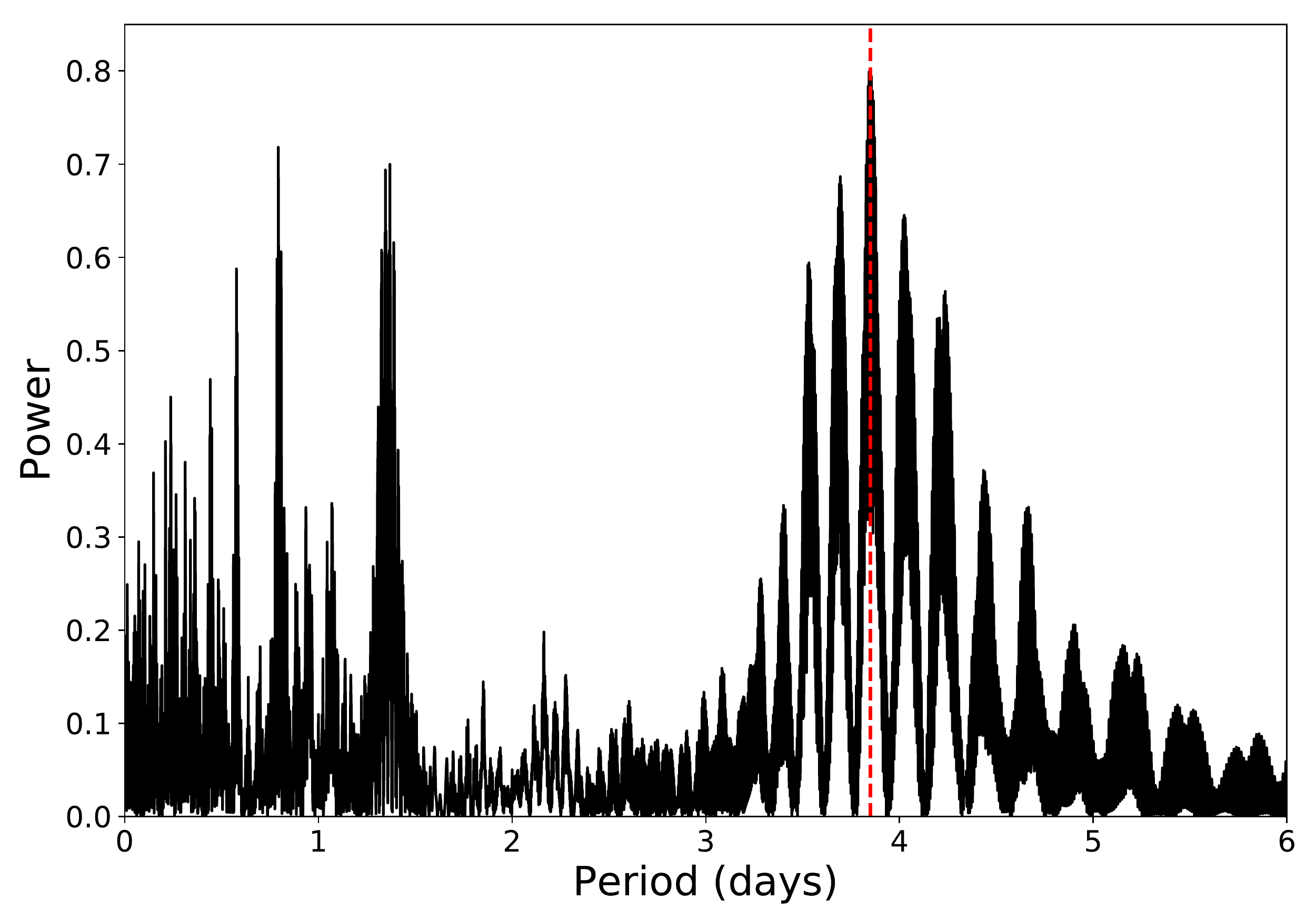}
\caption{Periodogram of the radial velocity data. The red dashed line marks the period recovered from the light curve. The peak with the same period as the transit light curve has a false alarm probability of 4.38 $\times$ \baseten{-5}.}
\label{fig:RVsignal}
\end{figure}

To account for the discrepancy between $v$ and \vsini, we could introduce a latitudinal differential rotation of ~20-40\%. According to \citet{cameron2007}, the magnitude of differential rotation for a star with \teff\ = 6216 K is estimated to be $\Delta\Omega$ $\sim$0.28 radian day$^{-1}$. From the rotation period of the star, the angular velocity of the spot is $\Omega$ $\sim$1.1 radian day$^{-1}$. The observed high \vsini\ could be explained if the spot that \kepler\ observed is long-lived, located at a relatively high latitude, and its angular velocity at the equator is $\sim$1.5 radian day$^{-1}$. 

\section{Orbital \& Planetary Parameters}

\subsection{Confirmation of Kepler-1658 b}

An unambiguous confirmation of a transiting exoplanet is typically performed by detecting radial velocity variations that are in phase with the ephemeris determined from transits. The initial three radial velocities (RVs) were taken as part of the Kepler Follow-up program during the \kepler\ mission. We obtained 20 more TRES RVs once we realized that it was possible that Kepler-1658 was hosting a planet (Figure \ref{fig:combined}), for a total of 23 RV observations.

As discussed in Section 3.3, Kepler-1658 is rotating rapidly, thus resulting in RVs with relatively large uncertainties. Despite this limitation, a periodogram of the RV data only reveals a highly significant peak that is fully consistent with the same period of the transit signal (Figure \ref{fig:RVsignal}). Additionally, phasing the RVs with the ephemeris and orbital period from \kepler\ reveals a clear variation with a semi-amplitude, \kstar\ = 579.45$^{+43.13}_{-42.94}$ \ms, well above the detection threshold set by the RV uncertainties (Figure \ref{fig:FinalFit}d). Therefore, the consistency between the RV and transit data unambiguously confirms Kepler-1658 b as a hot Jupiter.

\subsection{Transit \& RV Modeling}

\begin{table}[htb]
\begin{centering}
\caption{Model Parameters}
\begin{tabular}{L{5cm} C{3cm}}
\noalign{\smallskip}
\tableline\tableline
\noalign{\smallskip}
\textbf{Parameter} & \textbf{Prior} \\
\noalign{\smallskip}
\tableline\tableline
\noalign{\smallskip}
$z$ & $\mathcal{U}[-1;1]$ \\
$u_1\!^{\rm a}$ & $\mathcal{N}(0.3033;0.6)$\\
$u_2\!^{\rm a}$ & $\mathcal{N}(0.3133;0.6)$\\
 & $u_1$ $\ge$ 0 \\
 & $u_1$ + $u_2$ $\le$ 1 \\
 & $u_1$ + 2$u_2$ $\ge$ 0 \\
$\gamma$ (\ms) & $\mathcal{U}[-1200;-700]$ \\
K (\ms) & $\mathcal{U}[350;850]$ \\
$P$ (days)  & $\mathcal{U}[3.75;3.95]$ \\
$T_0$ (BKJD)$^{\rm b}$  & $\mathcal{U}[171.9;173.9]$ \\
$b$  & $\mathcal{U}[0;1]$ \\
$\rm R_{p}/R_\star$  & $\mathcal{U}[0.02;0.06]$ \\
$e\sin\omega$  & $\mathcal{U}[-1;1]$ \\
$e\cos\omega$  & $\mathcal{U}[-1;1]$ \\
$e$ & 1/$e$ \\
$\delta_{\text{occ}}$ (ppm) & $\mathcal{U}[0;200]$ \\
$\rho_{\star}$ (\gcc)$^{\rm c} $ & $\mathcal{N}(0.0834;0.0079)$ \\
\noalign{\smallskip}
\hline
\noalign{\smallskip}
\multicolumn{2}{l}{{\sc \textbf{Notes}} ---} \\
\multicolumn{2}{l}{$^{\rm a}$ Adopted from \citet{claret2011} archived tables}\\
\multicolumn{2}{l}{$\,\,\,$ with additional priors to prevent nonphysical values.} \\
\multicolumn{2}{l}{$^{\rm b}$ BKJD is the time system used by \kepler\ and is defined} \\
\multicolumn{2}{l}{$\,\,\,$ by Barycentric Julian Date (BJD) -- 2454833.} \\
\multicolumn{2}{l}{$^{\rm c}$ Adopted from the asteroseismic analysis.}\\
\noalign{\smallskip}
\end{tabular}
\end{centering}
\label{table:model}
\end{table}

\begin{figure*}
\centering
\includegraphics[width=\textwidth]{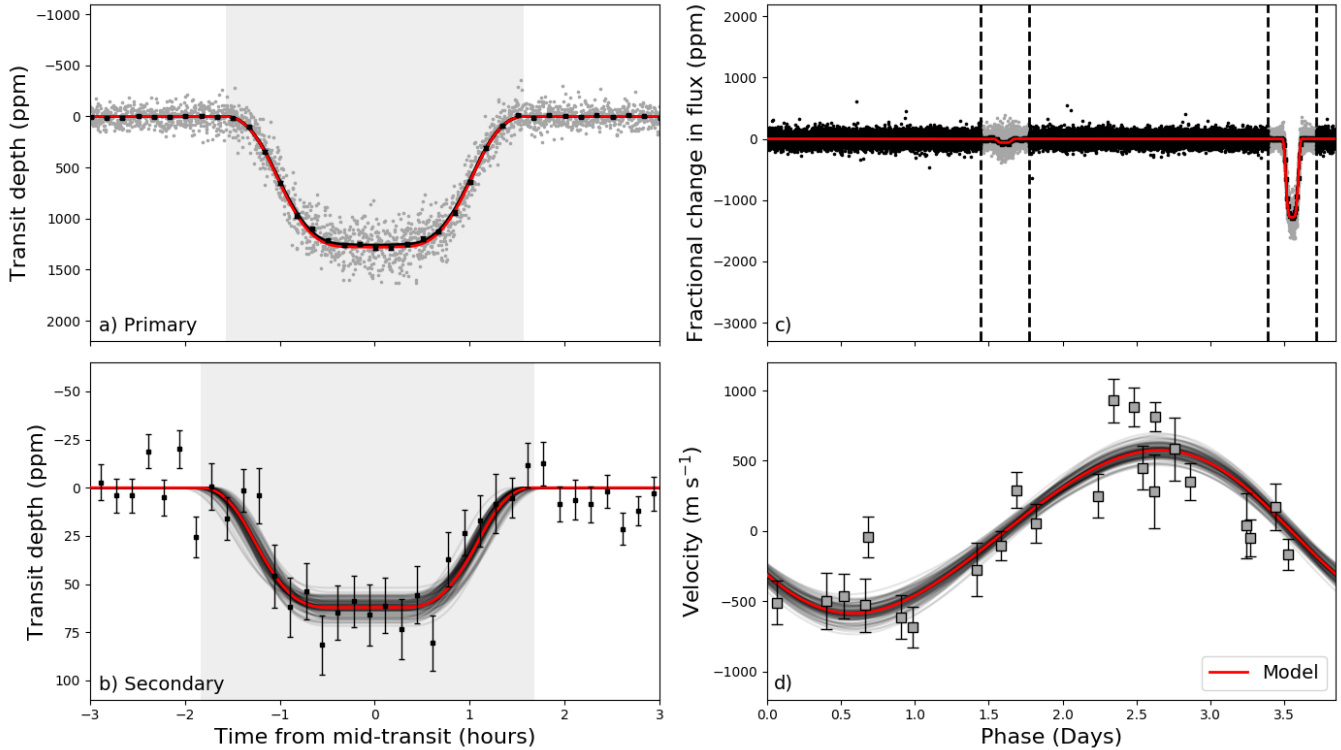}
\caption{Simultaneous transit and RV fit through MCMC analysis of Kepler-1658. Panels (a-b): Phase-folded light curve centered on the primary and secondary transits, where black squares are 10-min bins. The original data points from panel (b) have been removed for clarity. Panel (c): Full phase-folded light curve, where only gray points are used in the MCMC analysis. Panel (d): TRES RV observations. The median value for each parameter is used to generate a model and is shown in red. One hundred samples are drawn at random and added with transparency to give an idea of the uncertainty in the model.}
\label{fig:FinalFit}
\end{figure*}

To perform a combined transit and radial velocity fit we used the TRES measurements in Table \ref{tab:rvs} and \kepler\ long-cadence data, which cover a 4 times longer baseline than short-cadence data. For computational efficiency, we only used three times the transit duration centered at the time of mid-transit for both the primary and secondary eclipses (Figure \ref{fig:FinalFit}c).

\begin{figure*}
\centering
\includegraphics[width=\textwidth]{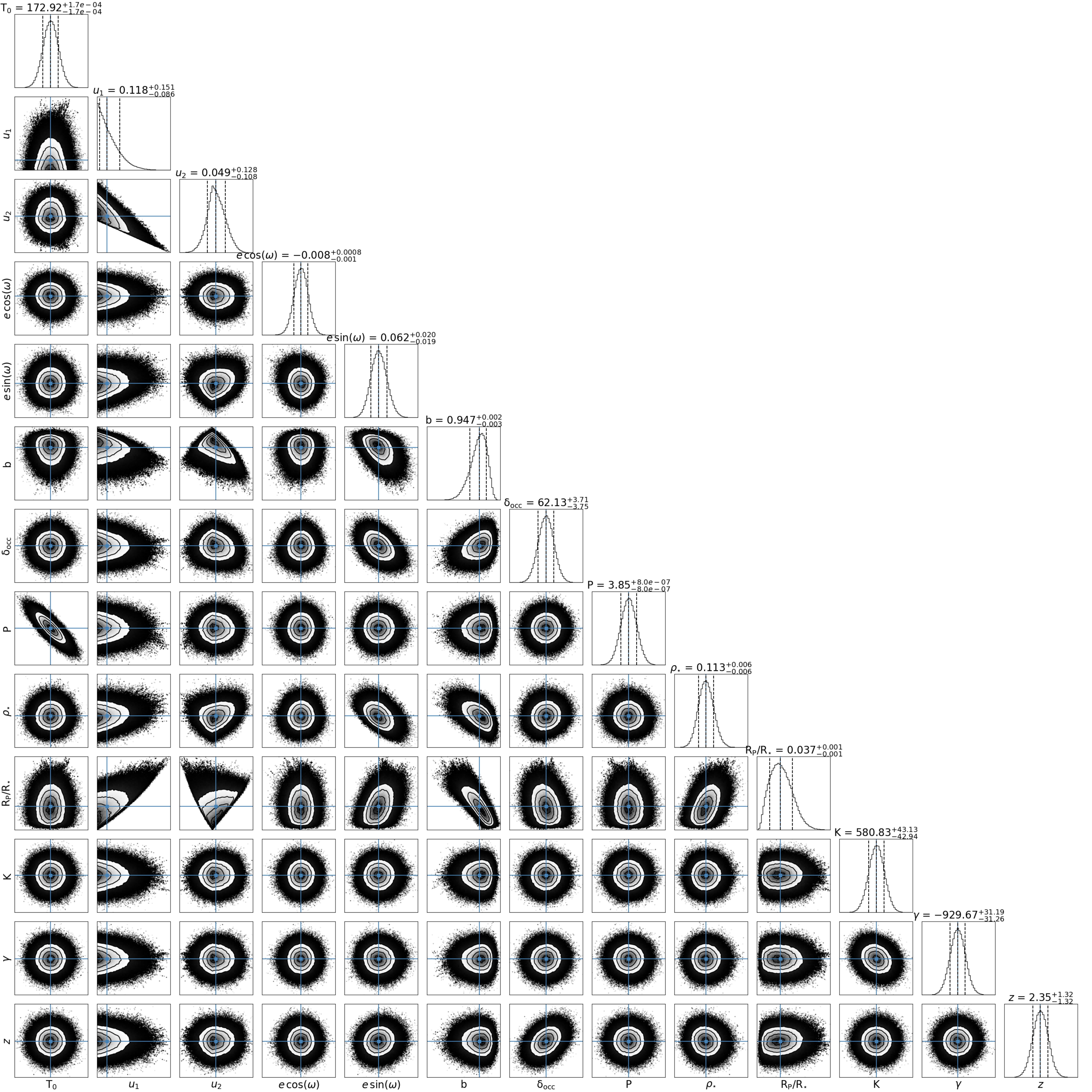}
\caption{Posterior distributions for 13-dimensional MCMC analysis of the simultaneous transit and RV fit. The parameters across the bottom are: time of mid-transit ($T_0$), linear ($u_1$) and quadratic ($u_2$) limb-darkening coefficients, eccentricity times the cosine of the argument of periastron ($e\cos\omega$), eccentricity times the sine of the argument of periastron ($e\sin\omega$), impact parameter (b), occultation depth (\occ), orbital period ($P$), mean stellar density (\rhostar), ratio of the planetary radius to the stellar radius (\rprs), velocity semi-amplitude (\kstar), systematic RV offset ($\gamma$), and photometric zero point ($z$).}
\label{fig:Corner}
\end{figure*}

We modeled the light curve and RV observations using \texttt{ktransit}\footnote[1]{\url{https://github.com/mrtommyb/ktransit}}, an implementation of the analytical model by \citet{mandel2002}. We assumed a linear ephemeris (constant orbital period) and quadratic limb darkening law.  The model fitted for the following parameters: orbital period ($P$), time of mid-transit ($T_0$), linear ($u_1$) and quadratic ($u_2$) limb-darkening coefficients, mean stellar density (\rhostar), systematic RV offset ($\gamma$), eccentricity times the sine of the argument of periastron ($e\sin\omega$), eccentricity times the cosine of the argument of periastron ($e\cos\omega$), occultation depth (\occ), impact parameter (b), ratio of the planetary radius to the stellar radius (\rprs), photometric zero point ($z$), and velocity semi-amplitude (\kstar). The predicted RV jitter due to stellar variability for a a star like Kepler-1658 is on the order of a few \ms\ \citep[e.g.][]{yu2018}, and thus negligible compared to the formal RV uncertainties ($\sim\, 100-200$\,\ms).

We used a Gaussian prior of 0.083 $\pm$ 0.008 \gcc\ for the mean stellar density derived from asteroseismology, as discussed in Section 3.2. Since Kepler-1658 is a rapid rotator resulting in large RV uncertainties, having an independent measurement of the mean stellar density helped break the degeneracies that exist between $e$ and $\omega$. Using stellar parameters \teff\ = 6216 K, \logg\ = 3.57 (dex), and [m/H] = -0.18, we extracted limb-darkening coefficients $u_1$ = 0.3033 and $u_2$ = 0.3133 from archived tables \citep{claret2011} by using the nearest grid point. We assigned Gaussian priors for the limb-darkening coefficients, using the \citet{claret2011} values as the center of the distribution with a width of 0.6. Additional priors on $u_1$ and $u_2$ are implemented, using linear combinations to prevent the parameters from taking nonphysical values \citep{burke2008,barclay2015}. We used a Jeffreys prior for eccentricity (1/$e$) to avoid a positive bias \citep{eastman2013}. The remaining parameters were assigned uniform (flat) priors and are listed in Table \ref{table:model}.

We explored the parameter space by fitting the planet transit and RV data simultaneously using \texttt{emcee}, a Markov chain Monte Carlo (MCMC) algorithm \citep{foreman-mackey2013}. We initialized 30 walkers, each taking \baseten{5} steps. A burn-in phase of 4$\times$\baseten{3} steps was removed from each chain before concatenating samples to obtain the final posterior distribution for each parameter. A corner plot is shown in Figure \ref{fig:Corner} to demonstrate convergence of the 13-dimensional parameter space, highlighting known correlations in the parameter space.

\subsection{System Parameters}

All physical parameters and corresponding uncertainties for Kepler-1658 derived from the joint transit and RV model can be found in Table \ref{tab:mcmc}. 
In addition to the transit and RV observations, asteroseismology was critical to break parameter degeneracies. Specifically, since Kepler-1658 b has such a short orbital period, an eccentricity, $e$ = 0.06 $\pm$ 0.02 initially seemed unlikely but is required due to the observed transit duration (2.6 hours) and the strong constraint from the asteroseismic mean stellar density. A visualization of this can be seen in Figure \ref{fig:FinalFit}(a-b), where the time from mid-transit of the primary and secondary is plotted on the same timescale. The difference in transit durations and the slight offset of the secondary eclipse can only be explained by a mild eccentricity since we have the strong prior constraint on the mean stellar density from asteroseismology.

\citet{furlan2017} reported planet radius correction factors (PRCF) for photometric contamination from nearby stellar companions. Depending on the separation and contrast ratio, the companion can contribute to the total flux throughout the phase of the orbit, including the primary transit, thus underestimating the radius of the planet. Since RV observations confirm the planet is orbiting the primary star, we obtained a final planetary radius of \radiusp\ = 1.07 $\pm$ 0.05 \radiusj\ using a PRCF = 1.0065\% $\pm$ 0.0003\% \citep{furlan2017}. We note that this is PRCF is based on a measured contrast in the LP600 filter (see Section \ref{sec:imaging}), which is commonly assumed to be similar to the \kepler\ bandpass \citep{law14,baranec2016,ziegler17}. The PRCF was taken into account before deriving physical planet parameters and is therefore taken into consideration for the final values listed in Table \ref{tab:mcmc}.

\begin{table}[ht]
\caption{MCMC Parameter Summary}
\centering
\begin{tabular}{L{1.7cm} C{1.25cm} C{1.25cm} C{1.25cm} C{1.25cm}}
\noalign{\smallskip}
\tableline\tableline
\noalign{\smallskip}
\textbf{Parameter} &  \textbf{Best-fit} & \textbf{Median} & \textbf{84\%} & \textbf{16\%} \\
\noalign{\smallskip}
\tableline\tableline
\noalign{\smallskip}
\multicolumn{5}{c}{Fitted Parameters} \\
\noalign{\smallskip}
\hline
\noalign{\smallskip}
$z$ (ppm) & 1.703 & 2.350 & +1.316 & -1.317 \\
$P$ (days) & 3.8494 & 3.8494 & +8.04-7 & -8.01e-7 \\
$T_0$ (BKJD) & 172.9241 & 172.9241 & +1.69-4 & -1.67e-4 \\
$b$ & 0.9501 & 0.9471 & +0.0025 & -0.0032 \\
$\rm R_{p}/R_\star$ & 0.0359 & 0.0369 & +0.0008 & -0.0007 \\
$e\sin\omega$ & 0.0580 & 0.0622 & +0.0198 & -0.0188 \\
$e\cos\omega$ & -0.0081 & -0.0084 & +0.0008 & -0.0008 \\
$\delta_{\text{occ}}$ (ppm) & 61.842 & 62.127 & +3.712 & -3.750 \\
$u_1$ & 0.0154 & 0.1179 & +0.1513 & -0.0859 \\
$u_2$ & 0.0453 & 0.0487 & +0.1276 & -0.1082 \\
$\gamma$ (\ms) & -915.88 & -929.67 & +31.19 & -31.26 \\
K (\ms) & 575.80 & 580.83 & +43.13 & -42.94 \\
$\rho_{\star}$ (\gcc) & 0.1068 & 0.1130 & +0.0063 & -0.0060 \\
\noalign{\smallskip}
\hline
\noalign{\smallskip}
\multicolumn{5}{c}{Derived Parameters} \\
\noalign{\smallskip}
\hline
\noalign{\smallskip}
$a$ (AU) & 0.0546 & 0.0544 & 0.0007 & -0.0007 \\
\radiusp\ (\radiusj)$^{\rm *}$ & 1.04 & 1.07 & +0.05 & -0.05 \\
$a/\text{R}_\star$ & 4.07 & 4.04 & 0.18 & -0.17 \\
$e$ & 0.0585 & 0.0628 & +0.0197 & -0.0185 \\
$\omega$ ($^{\text{o}}$) & -7.99 & -7.73 & +1.93 & -3.37 \\
\massp\ (\massj)$^{\rm *}$ & 5.88 & 5.88 & +0.47 & -0.46 \\
\rhop\ (\gcc) & 7.00 & 6.36 & +1.07 & -0.91 \\
$i$ ($^{\text{o}}$) & 76.55 & 76.52 & +0.58 & -0.59 \\
$\rm A_{\lambda}$ & 0.785 & 0.734 & +0.090 & -0.081 \\
\noalign{\smallskip}
\tableline
\noalign{\smallskip}
\multicolumn{5}{l}{$^{\rm *}$ Adopting a Jupiter radius of 6.9911 x \baseten{4} km and mass} \\
\multicolumn{5}{l}{$\,\,\,$ of 1.898 x \baseten{27} kg.} \\
\noalign{\smallskip}
\end{tabular}
\label{tab:mcmc}
\end{table}

The secondary eclipse allows an estimate of the planetary albedo. According to \citet{winn2010transits}, a geometric albedo can be determined by:
\begin{equation}
A_{\lambda} = \delta_{\text{occ}}(\lambda)\,\bigg(\frac{R_{\text{p}}}{a}\bigg)^{-2},
\end{equation}
where \occ\ is the occultation depth. Using Equation (2), we derived a geometric albedo of 0.724$^{+0.090}_{-0.081}$ in the \kepler\ bandpass. This value is consistent to within 1-2$\sigma$ of RoboVetter's analysis by \citet{coughlin2016}, which estimated a value of either 0.494$^{+0.186}_{-0.083}$ or 0.348$^{+0.145}_{-0.062}$, depending on the lightcurve detrending method used.

Following \citet{winn2010transits}, the planet inclination can be derived by: 
\begin{equation}
b_{\text{tra}} = \frac{a \cos i}{R_{\star}} \bigg(\frac{1 - e^2}{1 + e \sin \omega}\bigg),
\end{equation}
where $b_{\text{tra}}$ is the impact parameter $b$ of the primary transit and $a$ is the planet's semimajor axis. Using Equation (3), we obtained an inclination, $i$ =  76.52 $\pm$ 0.59\degrees. This is consistent with the short orbital period and high impact parameter, $b$ = 0.947 $\pm$ 0.003.

The inclination of the orbital plane of the planet is used to break the \massp $\,\sin i$ degeneracy that would otherwise exist from RV observations alone. Given that \massp\ $\ll$ \mstar, the data determine \massp/\mstar$^{2/3}$ through:
\begin{equation}
\frac{M_p}{(M_p + M_{\star})^{2/3}} = \frac{K_{\star}\sqrt{1 - e^2}}{\sin i} \bigg(\frac{P}{2 \pi G}\bigg)^{1/3},
\end{equation}
but not \massp\ itself \citep{winn2010transits}. Because asteroseismology provides a stellar mass, the planet mass can be determined through Equation (4), yielding \massp\ = 5.87 $\pm$ 0.46 \massj. Combined with the corrected radius, we find a bulk density of \rhop\ = 6.36$^{+1.07}_{-0.91}$ \gcc.

The mean stellar density modeled from the light curve, \rhostar\ = 0.113 $\pm$ 0.006, converged 3$\sigma$ higher than the mean stellar density determined through asteroseismology. Analytical solutions for transit models are based on assumptions that the scaled semi-major axis, $a/$\rstar\ $\gtrsim$ 8 and the impact parameter, $b \ll$ 1 \citep{seager2003,winn2010transits}. We speculate that this inconsistency is due to the fact that the Kepler-1658 system falls in a parameter space where these approximations no longer hold ($a/$\rstar\ $\approx$ 4, $b$ $\approx$ 0.95).

\section{Discussion}

\begin{figure*}
\centering
\includegraphics[width=\textwidth]{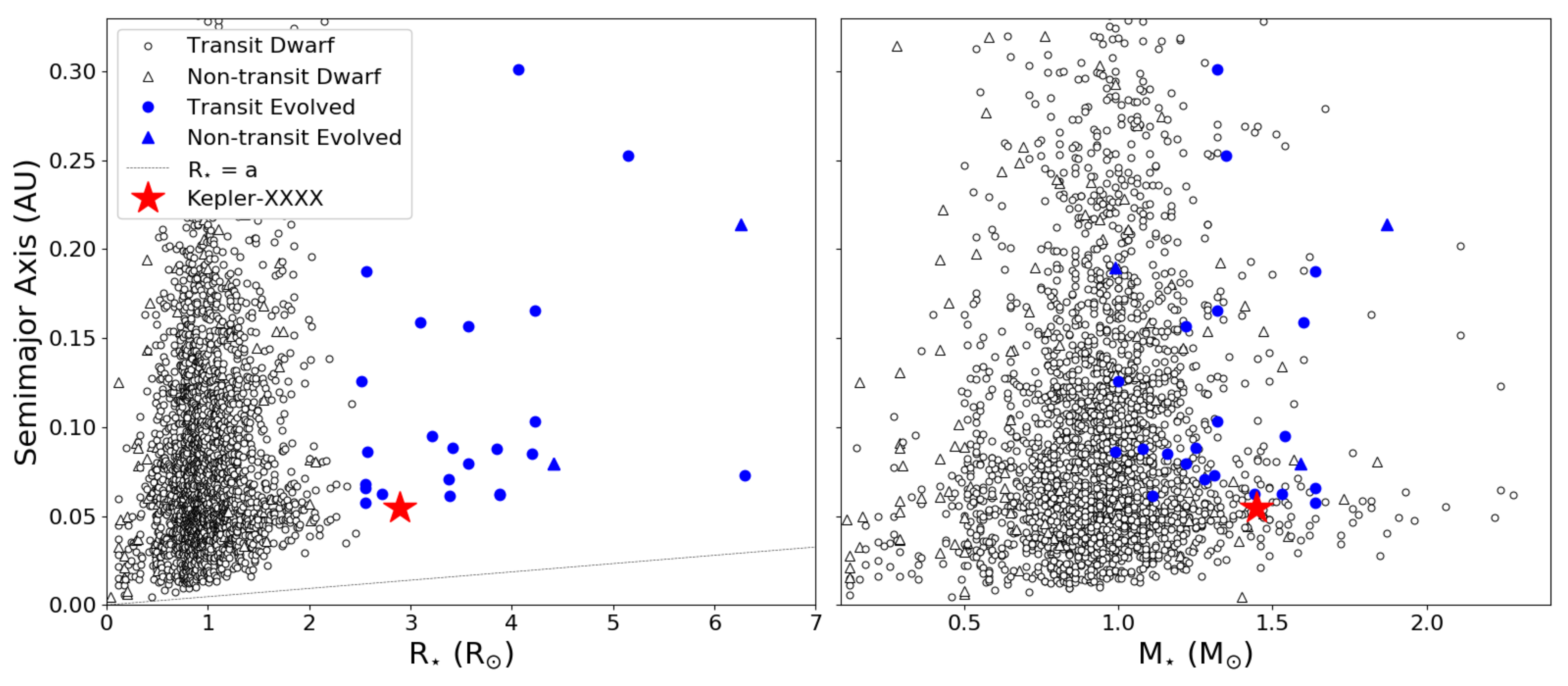}
\caption{Confirmed exoplanets taken from the NASA exoplanet archive (accessed on November 7, 2018, with the error bars omitted for clarity) Left: semimajor axis vs. stellar radius, where the dotted line represents \rstar $ = a$. Of all the evolved stars, Kepler-1658 is the closest short-period planet orbiting an evolved star. Right: semimajor axis vs. stellar mass. Short-period planets ($\le$ 100 days) orbiting evolved stars ($\ge$ 2.5 \rsun) are shown in blue, with values taken from: HD 102956 \citep{johnson2010}, Kepler-56 \citep{huber2013misalignment}, Kepler-278 and Kepler-391 \citep{rowe2014}, Kepler-91 \citep{lillo-box2014}, Kepler-432 \citep{ciceri2015,quinn2015}, Kepler-435 \citep{almenara2015}, K2-11 \citep{montet2015}, HIP 67851 \citep{jones2015}, 8 UMi \citep{lee2015}, K2-39 \citep{vaneylen2016}, K2-97 \citep{grunblatt2016}, Kepler-637, Kepler-815, Kepler-1004, and Kepler-1270 \citep{morton2016}, TYC 3667-1280-1 \citep{niedzielski2016}, K2-132 \citep{grunblatt2017}, HAT-P-67 \citep{zhou2017}, KELT-11 \citep{pepper2017}, WASP-73 \citep{stassun2017}, and 24 Boo \citep{takarada2018}.}
\label{fig:SMAvsStar}
\end{figure*}

\subsection{Orbital Period Decay} 

Kepler-1658 joins a rare population of exoplanets in close-in orbits around evolved, high-mass subgiant stars (Figure \ref{fig:SMAvsStar}). One theory for the lack of such planets is tidal evolution, occuring when a close-in giant planet tidally interacts with its host star \citep{levrard2009,schlaufman2013}. Tidal interactions in a 2-body system are  complex, and different driving mechanisms depend on a number of parameters \citep{zahn1989,barker2009,lai2012,rogers2013}. A system undergoing tidal dissipation conserves total angular momentum but dissipates its energy, and ultimately the dynamical evolution in the system is determined by the transfer of angular momentum between rotational and orbital parameters \citep{zahn1977,hut1981}.

\begin{figure*}
\centering
\includegraphics[width=\textwidth, height=11cm]{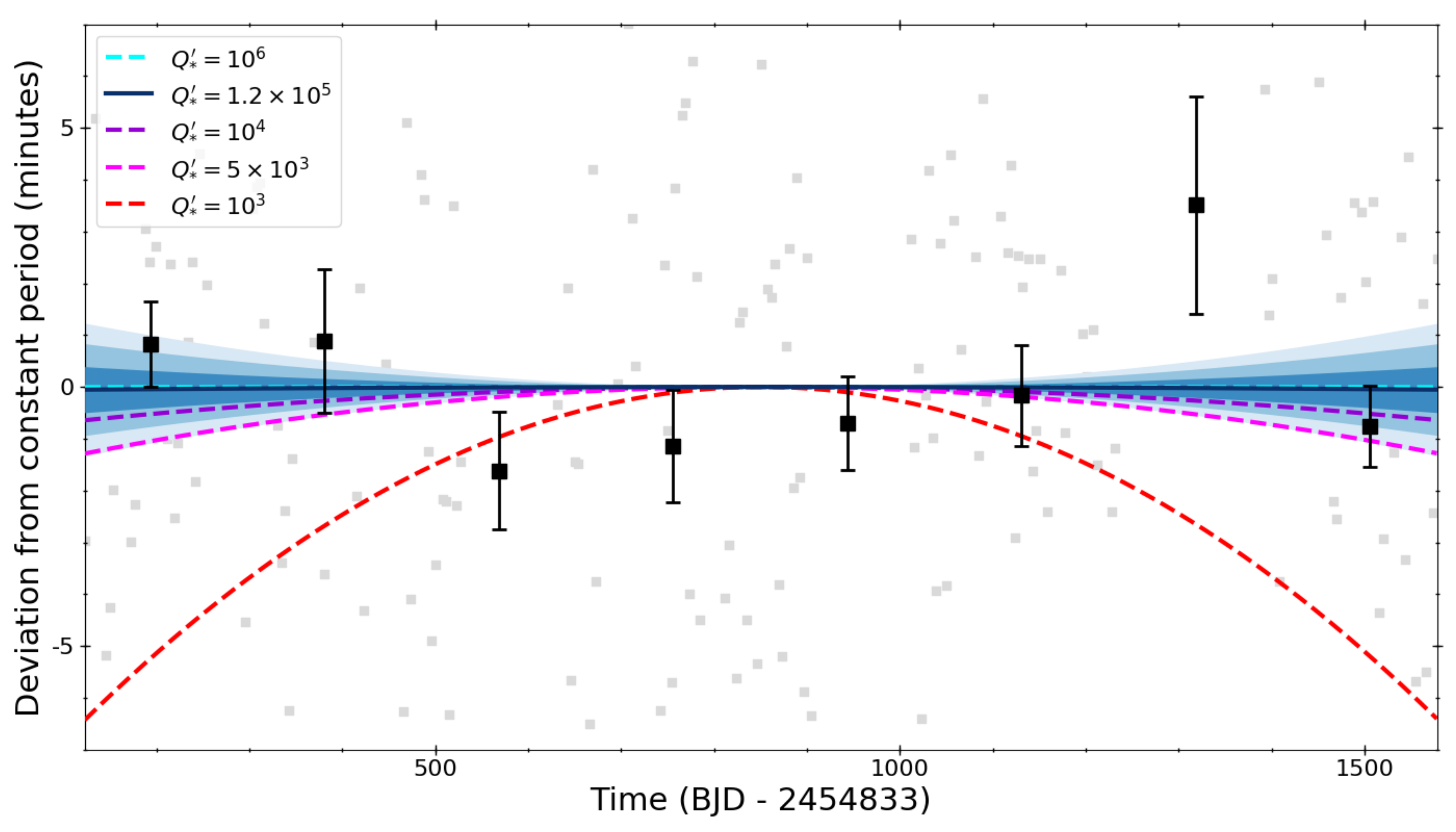}
\caption{Deviation from a constant orbital period versus epoch for Kepler-1658. Original transit times using \kepler\ long-cadence data are shown in gray and binned data is shown in black. Our analysis found a tidal quality factor of $Q_{\star}^{\prime}$ = 1.219 $\times$ \baseten{5}, shown by the blue line with 1$\sigma$, 2$\sigma$, and 3$\sigma$ levels shown in different transparencies. Theoretical values of $Q_{\star}^{\prime}$ = \baseten{3},\ 5 $\times$ \baseten{3}, \baseten{4}, and \baseten{6} are shown. Our results provide a strong lower limit for the tidal quality factor, ruling out $Q_{\star}^{\prime}$ $\le$ 4.826 $\times$ \baseten{3} for evolved subgiants for the first time observationally.}
\label{fig:OrbitalDecay}
\end{figure*}

There are only two possible outcomes to tidal evolution, depending on the stability of the system. One is a stable equilibrium in a coplanar, circular, synchronous orbit. If the total angular momentum in the system exceeds a critical value, then  the system becomes unstable, mostly dependent on the moments of inertia of both the host star and planet. A telltale sign of instability is when the orbital period of the planet is shorter than the rotational period of the star. When this happens, the planet deposits angular momentum onto the star, causing the star to spin up and the orbit to shrink \citep{levrard2009,adams2010,schlaufman2013,blecic2014,maciejewski2016,vaneylen2016}. In this latter scenario, there is no stable equilibrium point and the planet will migrate inwards until it is eventually engulfed by the host star. However, even systems that are marginally stable can be susceptible to inward planet migration due to evolutionary effects or angular momentum loss through magnetized winds \citep{vansaders2013}.

Following \citet{levrard2009}, we can estimate the timescale of orbital decay using:
\begin{equation}
\tau_a \simeq \frac{1}{48} \frac{Q'_{\star}}{n}\bigg(\frac{a}{R_{\star}}\bigg)^{5} \bigg(\frac{M_{\star}}{M_p}\bigg)
\end{equation}
where $n$ is the mean orbital angular velocity and $Q'_{\star}$ is the tidal quality factor, which is a single parameter that encapsulates physical processes that occur in tidal dissipation. A more recent paper by \citet{lai2012} suggests that $Q'_{\star}$ can vary for different tidal processes (e.g. orbital decay and spin-orbit alignment). In addition, tidal theory is very poorly constrained observationally, resulting in possible values for $Q'_{\star}$ that span several orders of magnitude, ranging from \baseten{2} to \baseten{10} \citep{levrard2009,barker2009,adams2010,schlaufman2013,blecic2014}. Therefore, tidal dissipation timescales remain highly uncertain. 
Observations of orbital period decay would provide better constraints on $Q'_{\star}$, which is currently poorly understood due to the complex nature of tidal interactions. 
Orbital period decay of a hot Jupiter was first proposed by \citet{lin1996} and was only recently detected in WASP-12b by \citet{maciejewski2016}. Based on ten years of transit data, \citet{maciejewski2016} reported a tidal quality factor of 2.5 $\times$ \baseten{5} for a main-sequence (MS) host star. \citet{patra2017} use new transit times of WASP-12b to further confirm evidence of period decay and find a consistent tidal quality factor of 2 $\times$ \baseten{5}.

For subgiant and giant stars such as Kepler-1658, \citet{schlaufman2013} suggest that the stars become more dissipative as they evolve off the MS, with $Q'_{\star}$ closer to \baseten{2}-\baseten{3}. Kepler-1658 is a prime target to constrain orbital period decay because it has a scaled semi-major axis of $a/R_\star \approx 4$ (Figure \ref{fig:SMAvsStar}) and the time scale of period decay is a sensitive function of this parameter (Equation 5). There is also evidence that the Kepler-1658 system is unstable because the orbital period of the planet is less than the stellar rotation period. 

\begin{table}[htb]
\begin{centering}
\caption{Individual Transit Times}
\begin{tabular}{C{3cm} C{2cm} C{2cm}}
\noalign{\smallskip}
\tableline\tableline
\noalign{\smallskip}
\textbf{Epoch (BJD)} & \textbf{84\%} & \textbf{16\%} \\
\noalign{\smallskip}
\tableline\tableline
\noalign{\smallskip}
2454955.88020 & 0.00176 & -0.00174 \\
2454959.72578 & 0.00121 & -0.00116 \\ 
2454967.43397 & 0.00112 & -0.00118 \\
2454971.28509 & 0.00158 & -0.00154 \\
2454975.12848 & 0.00131 & -0.00130 \\
2454978.97490 & 0.00132 & -0.00125 \\
2454982.82490 & 0.00173 & -0.00179 \\
2454986.67586 & 0.00152 & -0.00152 \\
2454990.52639 & 0.00138 & -0.00129 \\
2454994.37652 & 0.00196 & -0.00198 \\
2455005.92202 & 0.00189 & -0.00191 \\
2455009.77189 & 0.00129 & -0.00126 \\
2455013.62780 & 0.00333 & -0.00284 \\
\noalign{\smallskip}
\hline
\noalign{\smallskip}
\multicolumn{3}{l}{{\sc \textbf{Note}} ---} \\
\multicolumn{3}{l}{This table is published in its entirety on the journal}\\
\multicolumn{3}{l}{website in machine-readable format. A portion is shown} \\
\multicolumn{3}{l}{for formatting purposes. A version is also available in the} \\
\multicolumn{3}{l}{source materials.} \\
\noalign{\smallskip}
\end{tabular}
\end{centering}
\label{table:TTs}
\end{table}

In order to test for evidence of period decay, we divided the long-cadence light curves into segments with a length corresponding to the orbital period, P = 3.85 days. We modeled the individual transits using \texttt{ktransit} and used a similar MCMC analysis to determine the transit times. All other parameters were fixed to the values from the global fit discussed in Section 4.2. The individual transit times are shown in Figure \ref{fig:OrbitalDecay} and listed in Table \ref{table:TTs}. 

Following \citet{maciejewski2016}, we modeled the period decay rate by adding a quadratic term to the previously assumed linear ephemeris:

\begin{equation}
T_x = T_0 + Px + \frac{1}{2}\delta P x^2,
\end{equation}

where $x$ is the orbit number, $T_x$ is the time of mid-transit of $x$ orbit, and $\delta P$ is the change in orbital period between consecutive orbits ($P \dot{P}$). We used MCMC analysis described by \citet{foreman-mackey2013} to fit for $T_0$, $P$, and $\delta{P}$. We placed uniform priors on $T_0$ and $\delta{P}$ and a Gaussian prior on $P$, as derived from the combined transit and RV fit result. 

Our analysis found $\delta{P}$ = (-.2048 $\pm$ 1.723) $\times$ \baseten{-8} days, corresponding to a decay rate of $\dot{P}$ = (-16.8 $\pm$ 141.25) \msy, where a negative value is indicative of orbital decay. This corresponds to an infall timescale of 20 Myr. For consistency and comparison to the values reported in \citet{maciejewski2016}, we used:
\begin{equation}
Q'_{\star} = 9P\dot{P}^{-1}\,\frac{M_p}{M_{\star}}\bigg(\frac{R_{\star}}{a}\bigg)^5 \bigg(\omega_{\star} - \frac{2\pi}{P}\bigg)
\end{equation}
to estimate the tidal quality factor, where $\omega_{\star}$ is the stellar rotation rate. Using Equation 9, we calculated a tidal quality factor, \qstar\ = 1.219 $\times$ \baseten{5}, consistent with that reported by \citet{maciejewski2016}. Figure \ref{fig:OrbitalDecay} shows the timing residuals of a linear ephemeris plotted with the quadratic model, including our median fit with 1$\sigma$, 2$\sigma$, and 3$\sigma$ levels shown in different transparencies. Theoretical values are added for comparison.

Although the MCMC analysis is suggestive of period decay, the result is consistent with zero within 1$\sigma$, as seen in Figure \ref{fig:OrbitalDecay}. However, we can still provide a strong constraint for the tidal quality factor in subgiants. We report a 3$\sigma$ upper limit of $\delta{P}$ $\le$ -5.17 $\times$ \baseten{-8} days, or a decay rate $\dot{P}$ $\le$ -0.424 s yr$^{-1}$. This corresponds to a lower limit of $Q'_{\star}$ $\ge$ 4.826 $\times$ \baseten{3} and thus clearly rules out lower tidal quality factors suggested for subgiants in the literature and places strong constraints on tidal theories for evolved stars by effectively ruling out $\sim$2 orders of magnitude.

\subsection{Spin-orbit Misalignment}

Kepler-1658 is rapidly rotating, suggesting that it started farther up the main-sequence than the Sun, with a negligible or nonexistent convective envelope to effectively spin the star down through magnetic braking \citep{vansaders2013}. Solar-like oscillations are excited by near-surface convection, implying that host star now has a convection zone. More specifically, Kepler-1658 has \teff\ $\approx$ 6210 K, suggesting that it recently crossed the transition between exoplanet systems showing small and large spin-orbit misalignments  \citep{winn2010obliquities}. Therefore, Kepler-1658 may provide valuable insights into the dynamical formation history of hot Jupiters.

The obliquity, or the angle measured between the orbital angular momentum vector and rotational axis of the star, is defined as:
\begin{equation}
\cos\psi = \cos i_{\star} \cos i + \sin i_{\star} \sin i \cos \lambda
\end{equation}
where $\lambda$ is the sky-projected obliquity, $i$ is the planet's inclination, and $i_{\star}$ is the stellar inclination. The sky-projected obliquity $\lambda$ can be directly measured through the Rossiter-McLaughlin (RM) effect for close-in giant planets if that star is rotating fast enough \citep{rossiter1924,mclaughlin1924,ohta2005,winn2009}. The planet's inclination is trivially measured for transiting systems. 

The stellar inclination can be measured through relative amplitudes of rotationally split dipole modes \citep{gizon2003}. Stellar inclinations measured through asteroseismology demand a long baseline to achieve sufficient frequency resolution and SNR. The asteroseismic technique has been applied to more than a handful of  exoplanet hosts \citep{huber2013misalignment,chaplin2013inc,davies2015,campante2016misalign,kamiaka2018}, but is not applicable to Kepler-1658 due to the low S/N.

Alternatively, the stellar spin inclination can be measured if there is evidence of rotational modulation \citep{winn2007,schlaufman2010}. More specifically, the relation between the inclination, rotation period, and rotational velocity is given by:
\begin{equation}
\sin i_{\star} = \bigg(\frac{P_{\text{rot}}}{2 \pi R_{\star}}\bigg)\, v \sin i\,.
\end{equation}

The detection of the rotational period in Kepler-1658 allows us to put constraints on the stellar inclination through Equation (9). As discussed in Section 3.3, the observed \vsini\ and rotation period imply $i$ $\approx$ 90\degrees. With no prior information on the projected obliquity $\lambda$, the true obliquity can range  from 16.50\degrees\ $\le$ $\psi$ $\le$ 163.50\degrees, providing tentative evidence for a spin-orbit misalignment. Future spectroscopic observations while the planet is transiting would allow further constraints on the true obliquity. The expected signal of the RV anomaly due to the RM effect through:
\begin{equation}
\Delta v_{\text{RM}} \sim \text{\vsini}\, \bigg(\frac{R_p}{R_\star}\bigg)^2
\end{equation}
yielding $\Delta v_{\text{RM}}$ $\sim$ 55 \ms\ for KOI 4.01. The detection of the RM effect would add Kepler-1658 to the small number of systems for which true obliquity measurements are possible \citep{benomar2014,lund2014}.

\section{Conclusions}

We have used asteroseismology and spectroscopy to confirm Kepler-1658 b, \keplers\ first planet detection. Our main conclusions can be summarized as follows:

\begin{itemize}
\item Kepler-1658 is a subgiant with \teff\ = 6216 $\pm$ 78 K, \rstar\ = 2.89 $\pm$ 0.12 \rsun, and \mstar = 1.45 $\pm$ 0.06 \msun. As a massive subgiant, Kepler-1658 is currently undergoing a rapid phase of stellar evolution, joining only 9 known exoplanet hosts with similar properties (15 including statistically validated planets).

\item Kepler-1658 b is a hot Jupiter with \radiusp\ = 1.07 $\pm$ 0.05 \radiusj\ and \massp\ = 5.73 $\pm$ 0.45 \massj, with an orbital period of 3.85 days. The planet is part of a small population of short-period (P $\lesssim$ 100 days, a $\lesssim$ 0.5 AU) planets around evolved (\rstar\ $\gtrsim$ 2.5 \rsun, \logg\ $\lesssim$ 3.7) stars. Sitting at an orbital distance of only $\approx$ 0.05 AU, Kepler-1658 b is one of the closest known planets to an evolved star. We find tentative evidence for a mild eccentricity ($e$ = 0.06 $\pm$ 0.02), consistent with tidal evolution studies suggesting moderate eccentricities of short-period planets around evolved stars \citep{villaver2009,grunblatt2018}.

\item Individual transit times over 4 years of \kepler\ observations place a strong upper limit of the orbital period decay rate, $\dot{P}$ $\le$ -0.42 s yr$^{-1}$, setting a lower limit on the tidal quality factor $Q'_{\star}$ $\ge$ 4.826 $\times$ \baseten{3}. Our measurements provide the first strong observational limit on tidal quality factors of subgiant stars,  ruling out $\sim$2 orders of magnitude of suggested theoretical values in the literature. 

\item Kepler-1658 sits close to the proposed misalignment boundary of 6250 K for hot Jupiter obliquities. While the combination of rotation period, \vsini\ and stellar radius only provide tentative evidence for a high obliquity (16.50\degrees\ $\le$ $\psi$ $\le$ 163.50\degrees), future spectroscopic observations of the RM effect may be able to provide stronger constraints on obliquity damping and thus hot-Jupiter migration theories.
\end{itemize}

The \kepler\ field will be observed by the Transiting Exoplanet Survey Satellite \citep[TESS;][]{ricker2015} in mid-2019. Extending the baseline of transit observations to over a decade for Kepler-1658 will allow for a stronger constraint on orbital period decay in more evolved systems. Extrapolating our period decay analysis to the time Kepler-1658 would be observed by TESS would rule out another order of magnitude for the tidal quality factor in subgiant stars.

Kepler-1658 is typical of the asteroseimic host stars we expect to find with TESS. \citet{campante2016} estimated that TESS will find at least 100 asteroseismic exoplanet hosts, and the first detection by \citet{huber19} confirms that this yield with be biased towards evolved subgiants similar to Kepler-1658 due to the increase of asteroseismic detection probabilities with stellar luminosity. Since targets found by TESS will be more amenable to follow-up, we expect that larger samples of short-period planets around evolved stars will provide better clues into planet formation, tidal migration, and dynamical evolution studies.

\section{Acknowledgements}
The authors thank Eric Gaidos, Josh Winn, Tom Barclay, Jennifer van Saders, Lauren Weiss, Erik Petigura, Benjamin J. Fulton, Cole Campbell Johnson, Sam Grunblatt, Jamie Tayar, Jessica Stasik, and Connor Auge for helpful discussions. The authors wish to recognize and acknowledge the very significant cultural role and reverence that the summit of Maunakea has always had within the indigenous Hawai`ian community.  We are most fortunate to have the opportunity to conduct observations from this mountain.

A.C. acknowledges support by the National Science Foundation under the Graduate Research Fellowship Program. D.H. acknowledges support by the National Aeronautics and Space Administration (NNX14AB92G, NNX17AF76G, 80NSSC18K0362) and by the National Science Foundation (AST-1717000). D.W.L. acknowledges support from the \kepler\ mission under the National Aeronautics and Space Administration Cooperative Agreement (NNX13AABB58A) with the Smithsoniann Astrophysical Observatory. T.L.C. acknowledges support from the European Union's Horizon 2020 research and innovation programme under the Marie Sk\l{}odowska-Curie grant number 792848. T.H. acknowledges support by the Japan Society for Promotion of Science under KAKENHI grant number JP16K17660.

\facilities{Keck:I (HIRES),\ldots}

\software{
\texttt{astropy} \citep{astropy}, 
\texttt{emcee} \citep{emcee}, 
\texttt{Matplotlib} \citep{Matplotlib}, 
\texttt{Pandas} \citep{Pandas}, 
\texttt{SciPy} \citep{Scipy}
}

\bibliography{references}

\end{document}